\documentclass[english,twocolumn,10pt,Engish,superscriptaddress,nofootbible]{revtex4-1}
\usepackage[T1]{fontenc}
\usepackage[latin9]{inputenc}
\usepackage{color}
\usepackage{babel}
\usepackage{amsbsy}
\usepackage{amstext}
\usepackage{graphicx}
\usepackage{esint}
\usepackage[unicode=true,pdfusetitle,
 bookmarks=true,bookmarksnumbered=false,bookmarksopen=false,
 breaklinks=false,pdfborder={0 0 0},pdfborderstyle={},backref=false,colorlinks=true]
 {hyperref}
\hypersetup{
 pdfborderstyle=,linkcolor=blue,citecolor=cyan}

\makeatletter

\providecommand{\tabularnewline}{\\}

\usepackage{babel}

\usepackage{xcolor}

\makeatother

\begin{document}
\title{Light nuclei photoproduction in relativistic heavy ion ultraperipheral
collisions}
\author{Jin-Yu Hu}
\email{jinyuhu2000@mail.ustc.edu.cn}

\affiliation{Department of Modern Physics, University of Science and Technology
of China, Anhui 230026, China}
\author{Shuo Lin}
\email{linshuo@mail.ustc.edu.cn}

\affiliation{Department of Modern Physics, University of Science and Technology
of China, Anhui 230026, China}
\author{Shi Pu}
\email{shipu@ustc.edu.cn}

\affiliation{Department of Modern Physics, University of Science and Technology
of China, Anhui 230026, China}
\affiliation{Southern Center for Nuclear-Science Theory (SCNT), Institute of Modern
Physics, Chinese Academy of Sciences, Huizhou 516000, Guangdong Province,
China}
\author{Qun Wang}
\email{qunwang@ustc.edu.cn}

\affiliation{Department of Modern Physics, University of Science and Technology
of China, Anhui 230026, China}
\affiliation{School of Mechanics and Physics, Anhui University of Science and Technology,
Huainan,Anhui 232001, China}
\begin{abstract}
We have investigated light nuclei pair photoproduction in relativistic
heavy ion ultraperipheral collisions. As a first attempt, we employ
our previously developed quantum electrodynamics model, which incorporates
a wave-packet description of initial nuclei, to compute the cross
section for proton-antiproton pair photoproduction. The effective
vertex for the photon and proton interaction is chosen based on studies
of two-photon exchange effects in hadron physics. We present the transverse
momentum, invariant mass, and azimuthal angle distributions of proton-antiproton
pairs at $\sqrt{s_{NN}}=200$ GeV in Au+Au ultraperipheral collisions.
We observe a $\cos(2\phi)$ modulation and an almost negligible $\cos(4\phi)$
modulation in the azimuthal angle distribution. Our studies helps
us better understand the matter generated by light.
\end{abstract}
\maketitle

\section{Introduction}

In relativistic heavy ion collisions, two heavy nuclei are accelerated
to near the speed of light, collide, and generate extremely strong
electromagnetic fields on the order of $10^{17}-10^{18}$ Gauss \citep{Bzdak:2011yy,Deng:2012pc,Roy:2015coa}.
Although these fields decay rapidly in a vacuum, they can persist
for a long time after interacting with the medium generated by the
collisions \citep{Roy:2015kma,Pu:2016ayh,Pu:2016bxy,Siddique:2019gqh,Inghirami:2016iru,Zhang:2022lje}.
Such strong fields provide a novel platform to study quantum transport
phenomena (see Refs. \citep{Kharzeev:2020jxw,Gao:2020vbh,Hidaka:2022dmn}
for recent reviews), the Schwinger mechanism \citep{Schwinger:1951nm,Copinger:2018ftr,Copinger:2020nyx,Copinger:2022jgg},
and photon-nuclear and photon-photon interactions. In the early 20th
century, Fermi \citep{Fermi:1924tc} first suggested, and Williams
\citep{Williams:1934ad} and Weizs$\ddot{a}$cker \citep{vonWeizsacker:1934nji}
later improved, the idea that strong electromagnetic fields can be
regarded as a flux of quasi-real photons. In relativistic heavy ion
collisions, photon fluxes are significantly enhanced due to the large
nuclear charge number. Additionally, the virtuality of photons is
highly suppressed by the Lorentz factor, meaning the photons generated
by the moving nuclei can be considered real. Therefore, relativistic
heavy ion collisions offer a new opportunity to investigate photon-nuclear
and photon-photon interactions.

Photon-nuclear interactions in high-energy scenarios, known as the
photoproduction of vector mesons, have been recently measured in relativistic
heavy ion ultraperipheral collisions (UPCs), e.g. see the measurements
from RHIC-STAR \citep{STAR:2019wlg,STAR:2021wwq,STAR:2022wfe} and
LHC-ALICE experiments \citep{ALICE:2019tqa,ALICE:2020ugp,ALICE:2021gpt}.
In the photoproduction of vector mesons, either nucleus can emit a
photon while the other acts as a target, and vice versa. These two
possibilities combine at the amplitude level and produce an interference
effect, akin to a high-energy version of the double-slit interference
phenomenon. This interference phenomenon was recently observed in
UPCs \citep{STAR:2022wfe} as a significant breakthrough. Theoretical
studies have discussed this phenomenon using the color glass condensate
effective theory \citep{Xing:2020hwh,Hagiwara:2020juc,Hagiwara:2021xkf}
and the vector meson dominance model \citep{Zha:2020cst}. This phenomenon
has also been found to detect nuclear deformation effects in isobaric
collisions \citep{Mantysaari:2023prg,Lin:2024mnj}. Another interesting
photon-nuclear interaction is the photoproduction of di-jets, which
has been measured at LHC-CMS \citep{CMS:2022lbi} and discussed using
Wigner distributions \citep{Hagiwara:2017fye} or diffractive transverse
momentum-dependent parton distribution functions of gluons \citep{Iancu:2023lel}. 

The two key processes for photon-photon interactions in UPCs are light-by-light
scattering, measured by LHC-ATLAS \citep{ATLAS:2017fur}, and lepton
pair photoproduction, measured by RHIC-STAR \citep{STAR:2018ldd,STAR:2019wlg},
LHC-ATLAS \citep{ATLAS:2018pfw} and LHC-ALICE experiments \citep{ALICE:2022hvk}.
Since the photon generated by the moving nuclei is nearly real, as
mentioned, lepton pair photoproduction in relativistic heavy ion collisions
can also be considered as the Breit-Wheeler processes \citep{Breit:1934zz}.
Although early pioneering works proposed the equivalent photon approximation
\citep{Klein:2016yzr,Bertulani:1987tz,Bertulani:2005ru,Baltz:2007kq},
it has been found inadequate to accurately describe lepton pair photoproduction
in UPCs due to its lack of dependence on impact parameters and the
transverse momentum of the initial photons. Several theoretical methods
have been developed to address photon-photon interactions in UPCs,
including the generalized equivalent photon approximation and quantum
electrodynamics (QED) in the background field approach \citep{Wang:2022ihj,Vidovic:1992ik,Hencken:1994my,Hencken:2004td,Zha:2018tlq,Zha:2018ywo,Brandenburg:2020ozx,Brandenburg:2021lnj,Li:2019sin,Luo:2023syp},
methods based on the factorization theorem \citep{Klein:2018fmp,Klein:2020jom,Li:2019yzy,Xiao:2020ddm,Shi:2024gex},
and QED models incorporating a wave-packet description of nuclei \citep{Wang:2021kxm,Wang:2022gkd,Lin:2022flv}.
Higher-order effects, such as Coulomb corrections \citep{Ivanov:1998ka,Eichmann:1998eh,Segev:1997yz,Baltz:1998zb,Baltz:2001dp,Baltz:2003dy,Baltz:2009jk,Zha:2021jhf,Klein:2020jom,Sun:2020ygb}
and Sudakov effects \citep{Li:2019sin,Li:2019yzy,Klein:2018fmp,Klein:2020jom,Shao:2023zge,Shao:2022stc},
have also been considered. For further discussion, see the recent
review \citep{Shi:2023nko} and references therein.

The successful discovery of the Breit-Wheeler process in UPCs naturally
raises the question: could we observe the photoproduction of other
particle-antiparticle pairs in UPCs? In this work, we study light
nuclei pair photoproduction in UPCs using a QED model that incorporates
a wave-packet description of nuclei \citep{Wang:2021kxm,Wang:2022gkd,Lin:2022flv}.
As a first attempt, we compute the cross section for proton-antiproton
($p\bar{p}$) pairs, the lightest nuclei, at the Born level. The effective
photon-proton vertex is chosen following Refs. \citep{Guttmann:2011ze,Zhou:2009xb}.
We present results for the transverse momentum, invariant mass, and
azimuthal angle distributions of $p\bar{p}$ pairs. Recently, we also
notice the related work on $\gamma+\gamma\rightarrow p+\bar{p}$ through
a different schemes \citep{Zhang:2024mql} and $\gamma+\gamma\rightarrow\pi^{+}+\pi^{-}$
in $e^{+}+e^{-}$ collisions \citep{Jia:2024xzx}. 

The paper is organized as follows. In Sec. \ref{sec:Theoretical},
we briefly review our theoretical formula for the differential cross
section in UPCs. In Sec. \ref{sec:distriubtion}, we present the transverse
momentum, invariant mass, and azimuthal angle distributions for $p\bar{p}$
pairs photoproduction at $\sqrt{s_{NN}}=$200 GeV in UPCs. We summarize
our work and discuss the results in Sec. \ref{sec:Conclusion}.

\begin{figure}
\centering\includegraphics[scale=0.4]{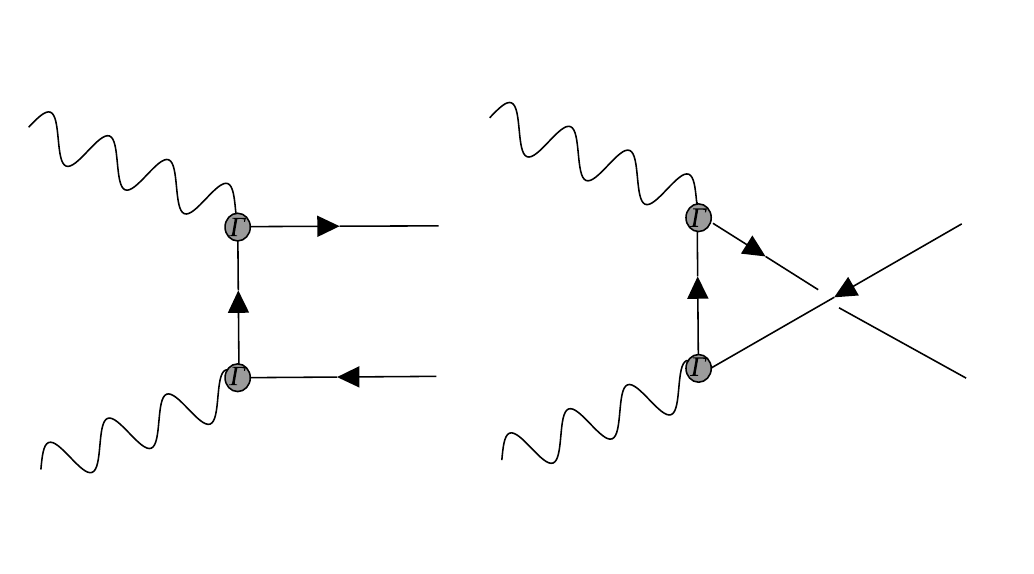}\caption{The Feynman diagram for the particle and anti-particle pairs in two
photon fusion processes at Born level. \label{fig:Pt-1}}
\end{figure}

\begin{figure*}[t]
\includegraphics[scale=0.2]{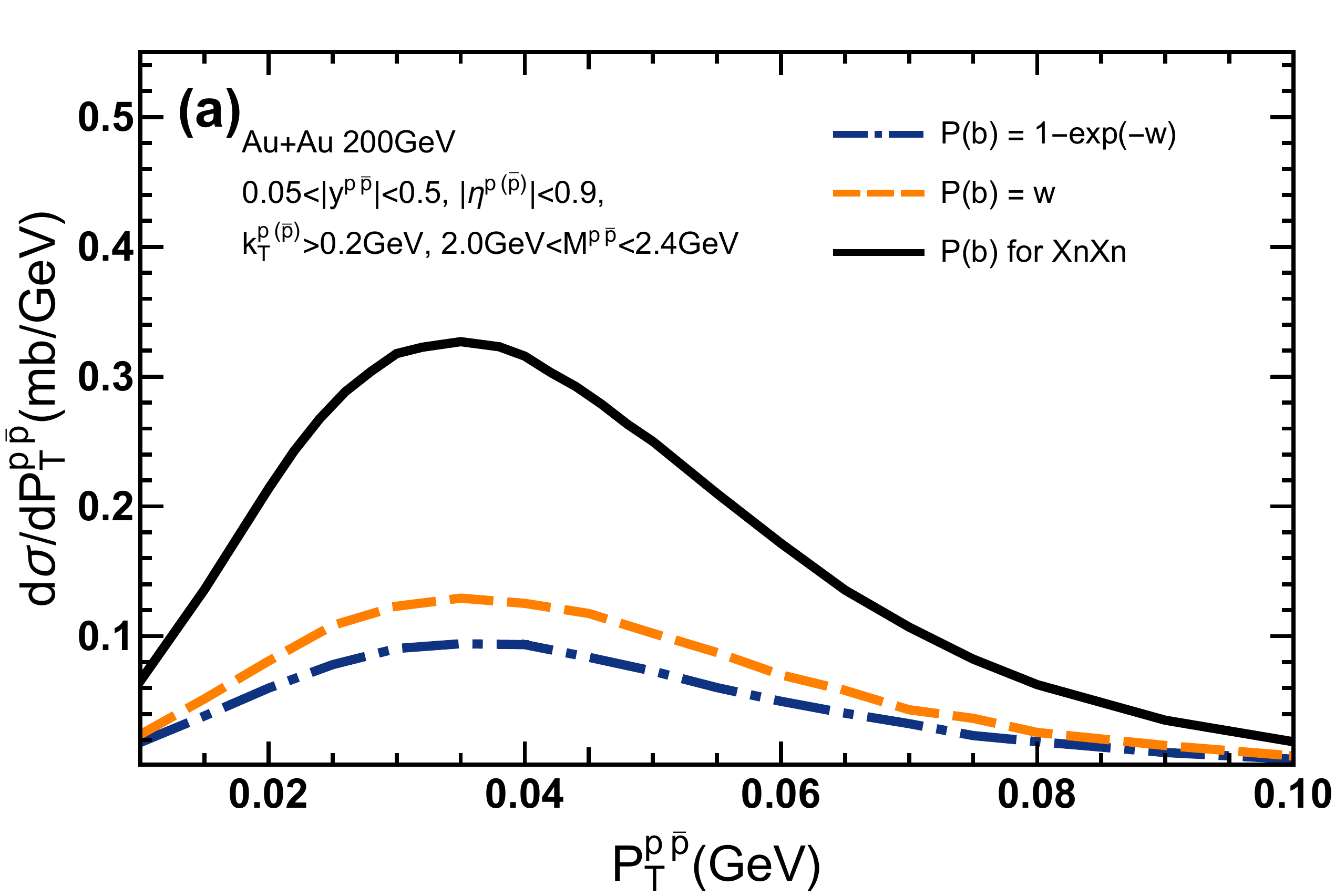}\includegraphics[scale=0.2]{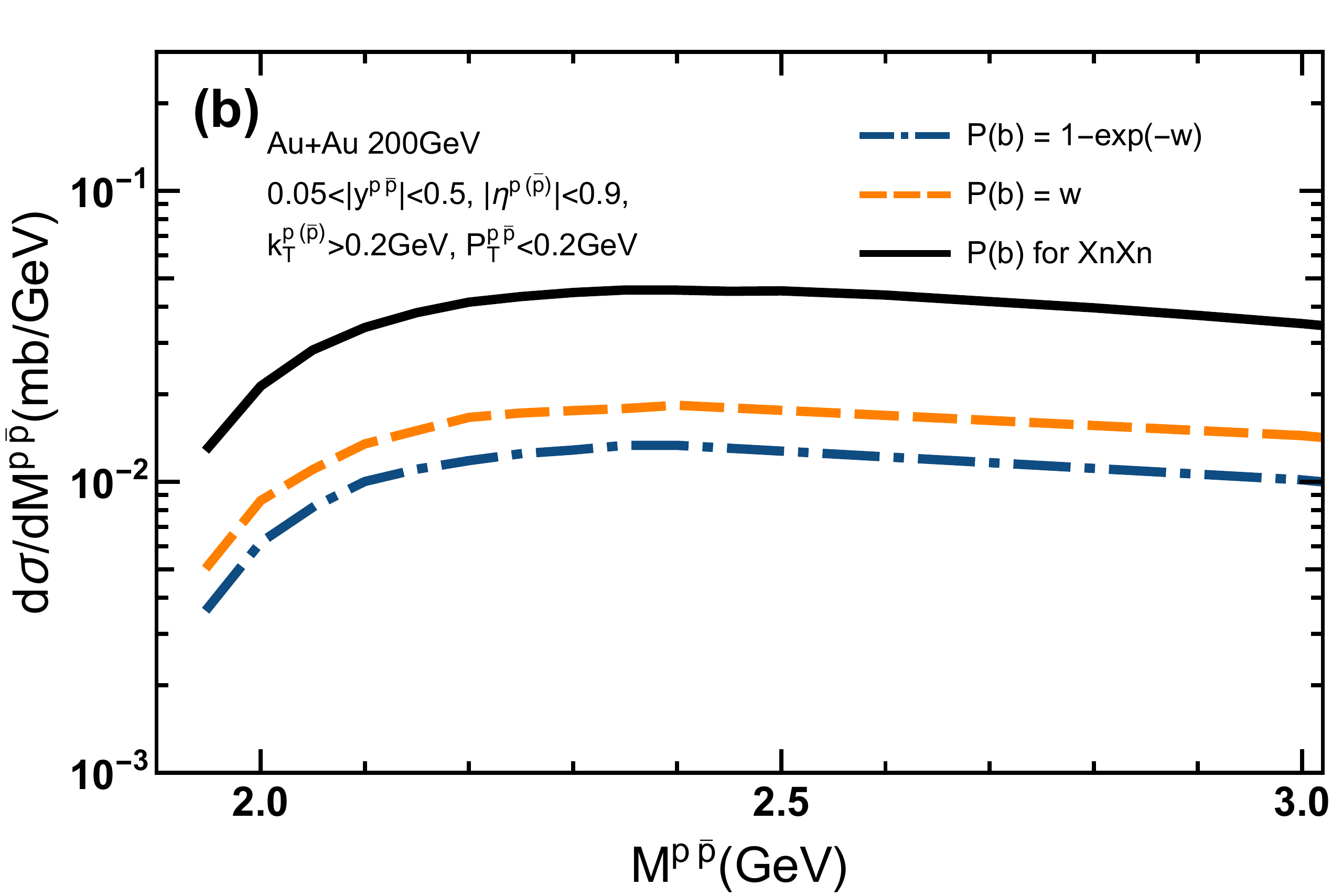}\caption{Differential cross sections as functions of (a) transverse momentum
$\mathbf{P}_{{\rm T}}^{p\overline{p}}$, (b) invariant mass $M^{p\bar{p}}$
for $p\bar{p}$ pairs. Different lines represent results computed
with different $\mathcal{P}(b_{T})$. The black solid lines use $\mathcal{P}(b_{T})$
from Eq. (\ref{eq:P_b_perp}) with $w$ from Eq. (\ref{eq:w_GDR}).
The orange dashed lines use $\mathcal{P}(b_{T})$ from Eq. (\ref{eq:P_b_temp})
with $w$ from Eq. (\ref{eq:w_GDR}). The blue dotted lines use $\mathcal{P}(b_{T})$
from Eq. (\ref{eq:P_b_temp}) with $w$ from Eq. (\ref{eq:w_GDR}).
\label{fig:Pt}}
\end{figure*}

\section{Theoretical framework \label{sec:Theoretical}}

We first review the theoretical framework for calculating the differential
cross section of particle and antiparticle pair photoproduction. The
framework follows Refs. \citep{Wang:2021kxm,Wang:2022gkd}, in which
some of us derived the differential cross section of lepton pair photoproduction
using the classical field approximation with a wave packet description
of nuclei. Consider two identical nuclei $A_{1}$ and $A_{2}$ move
in $\pm z$ direction with the velocity $u_{1,2}^{\mu}=\gamma(1,0,0,\pm v)$,
where $\gamma=1/\sqrt{1-v^{2}}$ is the Lorentz factor. We can use
a wave packet description for the initial nuclei. The sub-process
involves two photons from the colliding nuclei producing a particle-antiparticle
pair, described by
\begin{equation}
\gamma(p_{1})+\gamma(p_{2})\rightarrow H(k_{1})+\overline{H}(k_{2}),
\end{equation}
where $p_{1}^{\mu}$ and $p_{2}^{\mu}$ are four-momenta of photons,
and $k_{1}^{\mu}=(E_{k1},\mathbf{k}_{1})$ and $k_{2}^{\mu}=(E_{k2},\mathbf{k}_{2})$
are on-shell four-momenta for particle $H$ and anti-particle $\overline{H}$,
respectively. The Born-level total cross section can be expressed
in a compact form \citep{Wang:2021kxm,Wang:2022gkd},
\begin{eqnarray}
\sigma & = & \frac{Z^{4}e^{4}}{2\gamma^{4}v^{3}}\int d^{2}\mathbf{b}_{T}d^{2}\mathbf{b}_{1T}d^{2}\mathbf{b}_{2T}\delta^{(2)}(\mathbf{b}_{T}-\mathbf{b}_{1T}+\mathbf{b}_{2T})\nonumber \\
 & \times & \int\frac{d\omega_{1}d^{2}\mathbf{p}_{1T}}{(2\pi)^{3}}\frac{d\omega_{2}d^{2}\mathbf{p}_{2T}}{(2\pi)^{3}}\nonumber \\
 & \times & \int\frac{d^{2}\mathbf{p}_{1T}^{\prime}}{(2\pi)^{2}}e^{-i\mathbf{b}_{1T}\cdot(\mathbf{p}_{1T}^{\prime}-\mathbf{p}_{1T})}\frac{F^{*}(-\overline{p}_{1}^{\prime2})}{-\overline{p}_{1}^{\prime2}}\frac{F(-\overline{p}_{1}^{2})}{-\overline{p}_{1}^{2}}\nonumber \\
 & \times & \int\frac{d^{2}\mathbf{p}_{2T}^{\prime}}{(2\pi)^{2}}e^{-i\mathbf{b}_{2T}\cdot(\mathbf{p}_{2T}^{\prime}-\mathbf{p}_{2T})}\frac{F^{*}(-\overline{p}_{2}^{\prime2})}{-\overline{p}_{2}^{\prime2}}\frac{F(-\overline{p}_{2}^{2})}{-\overline{p}_{2}^{2}}\nonumber \\
 & \times & \int\frac{d^{3}k_{1}}{(2\pi)^{3}2E_{k1}}\frac{d^{3}k_{2}}{(2\pi)^{3}2E_{k2}}(2\pi)^{4}\delta^{(4)}(\overline{p}_{1}+\overline{p}_{2}-k_{1}-k_{2})\nonumber \\
 & \times & \sum_{\textrm{spin of }l,\overline{l}}\left[u_{1\mu}u_{2\nu}L^{\mu\nu}(\overline{p}_{1},\overline{p}_{2};k_{1},k_{2})\right]\nonumber \\
 & \times & \left[u_{1\sigma}u_{2\rho}L^{\sigma\rho*}(\overline{p}_{1}^{\prime},\overline{p}_{2}^{\prime};k_{1},k_{2})\right],\label{eq:cross section}
\end{eqnarray}
where $Z$ is the proton number of the nuclei, $\mathbf{b}_{iT}$
is the transverse position of the photon emission in nucleus $A_{i}$,
$\mathbf{b}_{T}$ is the impact parameter of colliding nuclei, $\overline{p}_{i}$
and $\overline{p}_{i}^{\prime}$ are photon momenta in the classical
field approximation and defined as ($i=1,2$ correspond to $+$ and
$-$), 
\begin{eqnarray}
\overline{p}_{i}^{\mu} & = & \left(\omega_{i},\boldsymbol{p}_{iT},\pm\frac{\omega_{i}}{v}\right),\quad\overline{p}_{i}^{\prime\mu}=\left(\omega_{i},\boldsymbol{p}_{iT}^{\prime},\pm\frac{\omega_{i}}{v}\right),
\end{eqnarray}
respectively. The form factor $F(-p^{2})$ is the Fourier transform
of the nuclear charge density distribution. In this work, we choose
the form factor as in Ref. \citep{Klein:2016yzr}
\begin{eqnarray}
F(p) & = & \frac{4\pi\rho^{0}}{p^{3}}[\sin(pR_{A})-pR_{A}\cos(pR_{A})]\frac{1}{a^{2}p^{2}+1},
\end{eqnarray}
where $a=0.7$ fm, $\rho^{0}=3A/(4\pi R_{A}^{3})$ is the normalization
factor, $A$ is the number of nucleons in the nucleus, $R_{A}=1.1A^{1/3}$
fm is the nucleus radius. 

The tensor $L^{\mu\nu}$ in Eq. (\ref{eq:cross section}) depends
on the final states generated through photoproduction. The typical
Feynman diagrams for particle-antiparticle pairs in two-photon fusion
processes at the Born level are presented in Fig. \ref{fig:Pt-1}.
If the particle $H$ is a fermion, then the effective $L^{\mu\nu}$
can be written in a compact form,
\begin{eqnarray}
 &  & L^{\mu\nu}(p_{1},p_{2};k_{1},k_{2})\nonumber \\
 & = & -\overline{u}(k_{1})[\Gamma^{\mu}S_{F}(k_{1},p_{1})\Gamma^{\nu}+\Gamma^{\nu}S_{F}(p_{1},k_{2})\Gamma^{\mu}]v(k_{2}),\nonumber \\
\label{eq: Lepton parts}
\end{eqnarray}
where $S_{F}$ is the effective propagator and $\Gamma^{\mu}$ is
the effective vertex, which will reduce to $\gamma^{\mu}$ in lepton
pair photoproduction \citep{Wang:2021kxm,Wang:2022gkd}.

Now, let us consider light nuclei photoproduction. As a first attempt,
we consider the photoproduction of the lightest nuclei and anti-nuclei
pair, i.e., $p\bar{p}$ pairs. Since light nuclei are not elementary
particles, the effective vertex $\Gamma^{\mu}$ and the effective
propagator $S_{F}$ must be chosen based on phenomenological models.
The photon-proton interactions are widely discussed in the context
of two-photon exchange effects in $e^{-}+p\rightarrow e^{-}+p$ and
$e^{+}+e^{-}\rightarrow p+\overline{p}$ \citep{Guttmann:2011ze,Zhou:2009xb,Blunden:2003sp,Guichon:2003qm},
also see recent review \citep{Carlson:2007sp,Arrington:2011dn} and
references therein. Several methods are also proposed to detect the
two two-photon exchange effects in experiments \citep{PANDA:2009yku,Nikolenko:2010zz,Kohl:2011zz}.

Therefore, here, we adopt the fermionic propagator $S_{F}(k,p)$ in
the lowest order of perturbative theory, 
\begin{equation}
-iS_{F}(k,p)=\frac{\gamma\cdot(k-p)+m_{p}}{(k-p)^{2}-m_{p}^{2}+i\varepsilon},
\end{equation}
with $m_{p}$ being the mass of proton and we follow Ref.\citep{Blunden:2003sp}
to adopt,
\begin{eqnarray}
\Gamma^{\mu}(q) & = & F_{1}(q^{2})\gamma^{\mu}+i\frac{F_{2}(q^{2})}{2m_{p}}\sigma^{\mu\nu}q_{\nu},\label{eq:vortex}
\end{eqnarray}
where $\sigma^{\mu\nu}=i[\gamma^{\mu},\gamma^{\nu}]/2$ and $F_{1}$
and $F_{2}$ are related to the Sachs form factors $G_{E}$ and $G_{M}$,
\begin{eqnarray}
F_{1}(q^{2}) & = & \frac{G_{E}(q^{2})+\tau G_{M}(q^{2})}{1+\tau},\nonumber \\
F_{2}(q^{2}) & = & \frac{G_{M}(q^{2})-G_{E}(q^{2})}{1+\tau},
\end{eqnarray}
with $\tau=-q^{2}/4m_{p}^{2}$. The form factors $G_{E}$ and $G_{M}$
are commonly parameterized as,
\begin{equation}
G_{E}(q^{2})=\frac{G_{M}(q^{2})}{\mu_{p}}=-\frac{\Lambda^{2}}{q^{2}-\Lambda^{2}},
\end{equation}
with $\mu_{p}$ being the magnetic moment of proton and $\Lambda=0.84$
GeV \citep{Blunden:2003sp}. In general, particle-antiparticle pairs
can also be generated through the $s$-channel. However, this contribution
may strongly depend on the effective coupling constant and is usually
found to be smaller than that of the current scheme (also see Ref.
\citep{Zhang:2024mql} for related discussions).

In RHIC-STAR experiments, the UPC events are triggered by detecting
the neutrons emitted from nuclei-photon interactions. To compare with
experimental data, we further need to add an extra factor $\mathcal{P}^{2}(b_{\perp})$
in the cross sections. The $\mathcal{P}(b_{\perp})$ is the probability
of emitting neutrons from the excited nucleus and can usually be parameterized
as \citep{Bertulani:1987tz},
\begin{equation}
\mathcal{P}(b_{T})=1-\exp(-w),\label{eq:P_b_perp}
\end{equation}
where $w$ is a function of $b_{T}$. When $w$ is small, $\mathcal{P}(b_{T})$
in Eq. (\ref{eq:P_b_perp}) is also approximately given by \citep{Hencken:2004td},
\begin{equation}
\mathcal{P}(b_{\perp})\simeq w.\label{eq:P_b_temp}
\end{equation}
One popular way to estimate $w$ is to use the Giant Dipole Resonance
model \citep{Bertulani:1987tz,Baur:1998ay,Hencken:2004td,Baltz:2009jk},
\begin{equation}
w=5.45\times10^{-5}Z^{3}(A-Z)/(A^{2/3}b_{T}^{2}).\label{eq:w_GDR}
\end{equation}
Since $w$ in Eq. (\ref{eq:w_GDR}) only depends on $A,Z$ and $b_{T}$,
the details of the nuclear mass distribution or deformation effects
are missing in this model. Another way to derive $w$ is to use the
equivalent photon approximation \citep{Brandenburg:2020ozx,Baltz:1998ex,Baltz:2002pp,Broz:2019kpl,Baltz:2009jk,Pshenichnov:2001qd},
\begin{equation}
w_{Xn}(b_{T})=\int d\omega n(\omega,b_{T})\sigma_{\gamma+A\rightarrow A^{\prime}+Xn}(\omega),\label{eq:XnXn_w}
\end{equation}
where the $X\geq1$ stands for the number of neutrons emitted by a
nucleus and $n(\omega,b_{T})$ is the photon flux \citep{Jackson:1998nia,Vidovic:1992ik,Wang:2021kxm},
and the photon-nucleus cross section $\sigma_{\gamma+A\rightarrow A^{\prime}+Xn}$
is given by fixed-target experiments \citep{Veyssiere:1970ztg,Berman:1987zz,Baltz:1998ex}.
In this work, we consider three kinds of $\mathcal{P}(b_{T})$, (i)
$\mathcal{P}(b_{T})$ from Eq. (\ref{eq:P_b_perp}) with $w$ from
Eq. (\ref{eq:w_GDR}), labeled as ``$\mathcal{P}(b_{T})=1-\exp(-w)$'',
(ii) $\mathcal{P}(b_{T})$ from Eq. (\ref{eq:P_b_temp}) with $w$
from Eq. (\ref{eq:w_GDR}), labeled as ``$\mathcal{P}(b_{T})=w$''
, (iii) $\mathcal{P}(b_{T})$ from Eq. (\ref{eq:P_b_perp}) with $w$
from Eq. (\ref{eq:XnXn_w}), labeled as ``XnXn''. These labels are
used in the following figures.

\begin{table}
\caption{The total cross section for $p\bar{p}$ pair photoproduction with
different choices of $\mathcal{P}(b_{T})$ at $\sqrt{s_{NN}}=200$
GeV and $54$ GeV in Au+Au UPCs. We have chosen $\mathbf{k}_{{\rm 1T}},\mathbf{k}_{{\rm 2T}}\protect\geq200$
MeV, $|y^{p\bar{p}}|\in[0.05,0.5]$ and $\eta^{p\bar{p}}\in[-0.9,-0.9]$.
The errors from the numerical integration are estimated by ZMCintegral
package.\label{tab:The-total-cross}}

\begin{tabular}{c|c|c|c}
\hline 
$\sqrt{s_{NN}}$ & $\mathcal{P}(b_{T})=1-e^{-w}$ & $\mathcal{P}(b_{T})=w$ & $\mathcal{P}(b_{T})$ for XnXn\tabularnewline
\hline 
\hline 
$200$ GeV & $4.5\pm0.1\mu\text{b}$ & $6.3\pm0.1\mu\text{b}$ & $15.7\pm0.3\mu\text{b}$\tabularnewline
\hline 
$54$ GeV & $51\pm3\text{nb}$ & $78\pm4\text{nb}$ & $170\pm1\text{nb}$\tabularnewline
\hline 
\end{tabular}
\end{table}

\section{Transverse momentum, invariant mass spectrum and azimuthal angle
distributions \label{sec:distriubtion}}

In this section, we present numerical results for the transverse momentum,
invariant mass, and azimuthal angle distributions of $p\bar{p}$ pair
photoproduction in Au+Au UPCs.  We define $\mathbf{P}_{{\rm T}}^{p\overline{p}}=\mathbf{k}_{{\rm 1T}}+\mathbf{k}_{{\rm 2T}}$
as the transverse momentum of the $p\bar{p}$ pair, where $\mathbf{k}_{{\rm 1T}}$
and $\mathbf{k}_{{\rm 2T}}$ are the transverse momentum for $p$
and $\bar{p}$, respectively, as introduced in Eq. (\ref{eq:cross section}).
Additionally, we also define $M^{p\overline{p}},\phi,$ and $y^{p\bar{p}}$
as the invariant mass of the $p\bar{p}$ pair, the angle between $\mathbf{P}_{{\rm T}}^{\mathrm{p\overline{p}}}$
and $\frac{1}{2}(\mathbf{k}_{{\rm 2T}}-\mathbf{k}_{{\rm 1T}})$, the
rapidity of $p\bar{p}$ pair. We use $\eta^{p(\bar{p})}$ for pseudo-rapidity
of single $p$ or $\bar{p}$. We set $\mathbf{k}_{{\rm 1T}},\mathbf{k}_{{\rm 2T}}\geq200$
MeV, $|y^{p\bar{p}}|\in[0.05,0.5]$ and $\eta^{p(\bar{p})}\in[-0.9,0.9]$.
The high dimensional integration in Eq.~(\ref{eq:cross section})
is computed by ZMCintegral package \citep{Wu:2019tsf,Zhang:2019nhd}. 

We first present the total cross section for $p\bar{p}$ pairs photoproduction
at the $\sqrt{s_{NN}}=200$ GeV and $54$ GeV in Au+Au UPCs with the
different choice of $\mathcal{P}(b_{T})$ in Table \ref{tab:The-total-cross}.
We observe that the total cross section are consistently in the range
of $\sim4-10\mu\textrm{b}$ at $\sqrt{s_{NN}}=200$ GeV and $\sim50-170$
nb at $\sqrt{s_{NN}}=54$ GeV, respectively. 

The differential cross section as a function of transverse momentum
$\mathbf{P}_{{\rm T}}^{\mathrm{p\overline{p}}}$ for $p\bar{p}$ pairs
is shown in Fig. \ref{fig:Pt}(a). We observe that the $d\sigma/d\mathbf{P}_{{\rm T}}^{p\overline{p}}$
ranges from $0.1$ to $0.3$ $\textrm{mb}/\textrm{GeV}$, with peaks
around $0.035-0.040$ GeV. The differential cross section with $\mathcal{P}(b_{T})$
from XnXn is slightly higher than those from the other two forms of
$\mathcal{P}(b_{T})$. 

We also plot the invariant mass $M^{p\overline{p}}$ distribution
of $p\bar{p}$ pairs in Fig. \ref{fig:Pt}(b). Interestingly, the
differential cross section grows smoothly when $M^{p\overline{p}}<2.4$
GeV, which contrasts with the rapid rise near the threshold seen in
lepton pair photoproduction. This difference arises from the non-trivial
effective electromagnetic form factors for protons in Eq. (\ref{eq:vortex}).
For $M^{p\overline{p}}>2.4$ GeV, the differential cross section decreases
slowly. Therefore, further experimental measurements of $p\bar{p}$
pairs photoproduction can provide additional constraints on the effective
vertex $\Gamma^{\mu}$.

\begin{figure}[t]
\centering\includegraphics[scale=0.23]{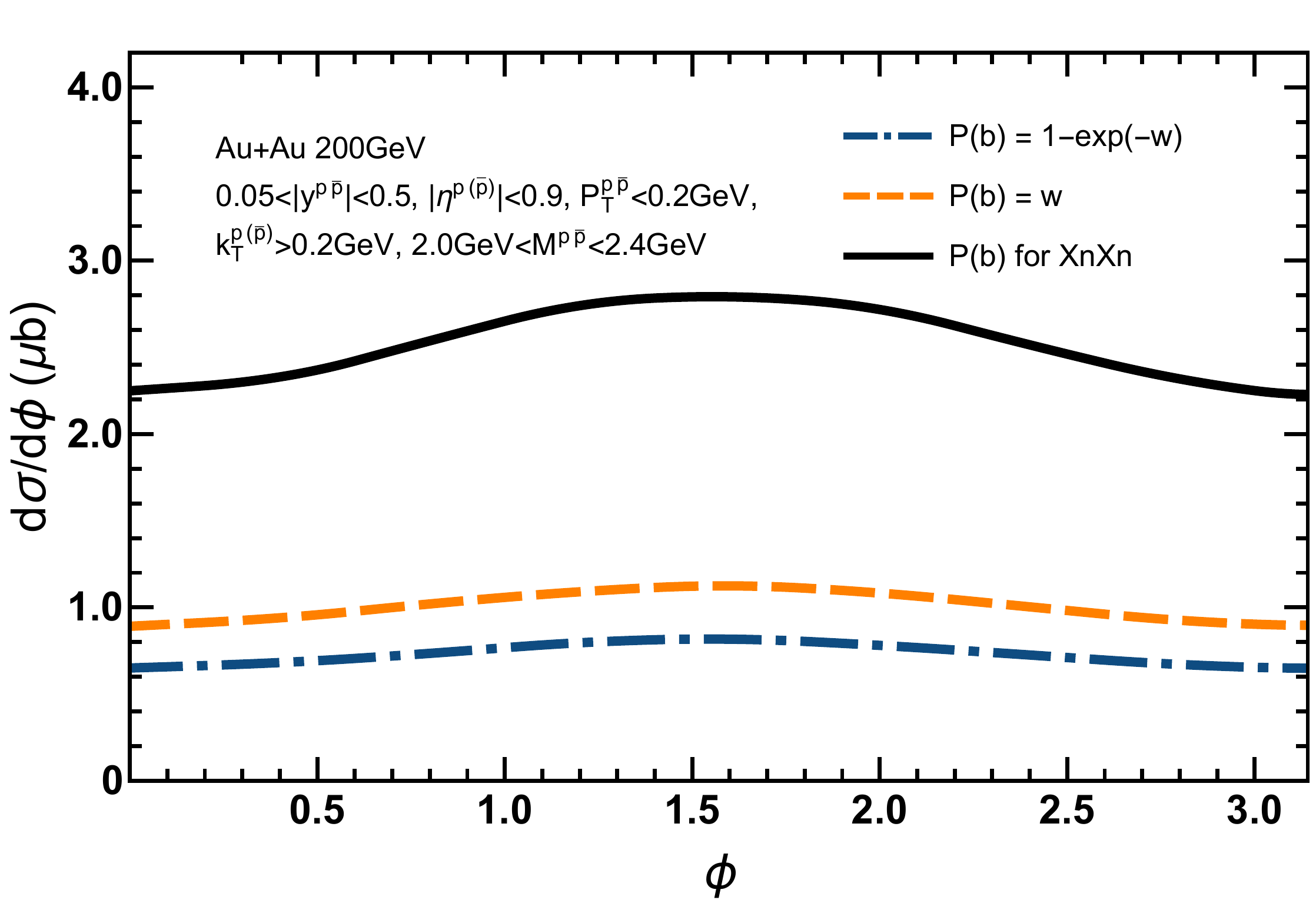}\caption{Differential cross sections as a function of azimuthal angle $\phi$.
The different lines have the same meanings as in Fig. \ref{fig:Pt}.
\label{fig:cos}}
\end{figure}

Finally, we plot the azimuthal angle distribution for $p\bar{p}$
pairs in Fig. \ref{fig:cos}. The azimuthal angle distribution for
lepton pair production exhibits $\cos(4\phi)$ and $\cos(2\phi)$
modulation due to the linear polarization of initial photons \citep{Li:2019sin,Li:2019yzy}.
We fit the differential cross section $d\sigma/d\phi$ using the function
$\mathcal{A}_{0}+\mathcal{A}_{2}\cos(2\phi)+\mathcal{A}_{4}\cos(4\phi)$.
The coefficients $\mathcal{A}_{0},\mathcal{A}_{2}$ and $\mathcal{A}_{4}$
are shown in Table \ref{tab:modulation}. In previous studies of lepton
pair photoproduction \citep{Li:2019sin,Li:2019yzy,Wang:2022gkd},
it was found that the $\cos(2\phi)$ modulation is proportional to
the mass squared of leptons. It was observed that dimuon pair photoproduction
has a much larger $\cos(2\phi)$ modulation than dielectron production.
Therefore, it is not surprising to observe $\cos(2\phi)$ modulation
for $p\bar{p}$ pairs photoproduction, despite the significant differences
in the vertex $\Gamma^{\mu}$ for protons in Eq. (\ref{eq:vortex})
compared to leptons. Interestingly, we do not observe significant
$\cos(4\phi)$ modulation in Fig. \ref{fig:cos}, and $\mathcal{A}_{4}$
is always an order of magnitude smaller than $\mathcal{A}_{2}$ in
Table \ref{tab:modulation}. Therefore, measurements of the azimuthal
angle can also impose constraints on the vertex $\Gamma^{\mu}$.

\begin{table}
\caption{We fit the $d\sigma/d\phi$ using the function $\mathcal{A}_{0}+\mathcal{A}_{2}\cos(2\phi)+\mathcal{A}_{4}\cos(4\phi)$.
The coefficients $\mathcal{A}_{0},\mathcal{A}_{2}$ and $\mathcal{A}_{4}$
for different $\mathcal{P}(b_{T})$. \label{tab:modulation}}

\begin{tabular}{c|c|c|c}
\hline 
 & $\quad\mathcal{A}_{0}(\mu\textrm{b})\quad$ & $\quad\mathcal{A}_{2}(\mu\textrm{b})\quad$ & $\quad\mathcal{A}_{4}(\mu\textrm{b})\quad$\tabularnewline
\hline 
\hline 
$\mathcal{P}(b_{T})=1-e^{-w}$ & $0.73$ & $-0.08$ & $\sim10^{-3}$\tabularnewline
\hline 
$\mathcal{P}(b_{T})=w$ & $1.01$ & $-0.11$ & $\sim10^{-3}$\tabularnewline
\hline 
$\mathcal{P}(b_{T})$ for XnXn & $2.52$ & $-0.28$ & $0.01$\tabularnewline
\hline 
\end{tabular}
\end{table}

\section{Conclusion and discussion \label{sec:Conclusion}}

In this work, we studied light nuclei photoproduction in UPCs using
our QED model that incorporates a wave-packet description of nuclei.
As a first attempt, we considered the photoproduction of the lightest
nuclei pairs, $p\bar{p}$ pairs, at the Born level in UPCs. The effective
vertex between photons and protons, $\Gamma^{\mu}$ , was chosen based
on studies of two-photon exchange effects, including information on
the electromagnetic form factors of the proton. We then present the
transverse momentum and invariant mass distributions of $p\bar{p}$
pairs at $\sqrt{s_{NN}}=200$ GeV in UPCs in \ref{fig:Pt}(a) and
(b), respectively. Interestingly, we observe a significant $\cos(2\phi)$
modulation and an almost vanishing $\cos(4\phi)$ modulation in the
azimuthal angle distribution of $p\bar{p}$ pairs in Fig. \ref{fig:cos}.
Our study is closely linked to two-photon exchange effects in hadron
physics. Therefore, further measurements in UPCs can also provide
constraints on these effects.

Light nuclei photoproduction opens a new window to generate light
nuclei and helps us better understand the matter generated by light.
Following the similar strategy, one can also consider deuterium or
helium nuclei photoproduction within our theoretical framework. For
deuterium nuclei, which are spin-$1$, the propagator in tensor $L^{\mu\nu}$
needs to be replaced accordingly, and the effective vertex may include
contributions from the electric quadrupole moment. For helium nuclei,
more complex nuclear information may be involved. Further discussions
along this line will be presented elsewhere.
\begin{acknowledgments}
We would like to thank Wang-Mei Zha, Xin Wu and Cheng Zhang for helpful
discussion. This work is supported in part by the National Key Research
and Development Program of China under Contract No. 2022YFA1605500,
by the Chinese Academy of Sciences (CAS) under Grants No. YSBR-088
and by National Nature Science Foundation of China (NSFC) under Grants
No. 12075235, No. 12135011, No. 12275082, No. 12035006, and No. 12075085.
\end{acknowledgments}

\bibliographystyle{h-physrev}
\bibliography{1}

\end{document}